\documentclass[reprint,amsmath,amssymb,aps]{revtex4-2}
\usepackage{graphicx,siunitx}       
\usepackage{dcolumn, booktabs}      
\usepackage{bm,lipsum,tabularx}    

\usepackage{xcolor,ifthen}
\newboolean{showred}
\newboolean{showblue}
\setboolean{showred}{false}   
\setboolean{showblue}{false} 
\newcommand{\highlightred}[1]{%
  \ifthenelse{\boolean{showred}}{\textcolor{red}{#1}}{#1}}
\newcommand{\highlightblue}[1]{%
  \ifthenelse{\boolean{showblue}}{\textcolor{blue}{#1}}{#1}}

\usepackage[pdfpagelabels, pdfencoding=auto, psdextra]{hyperref}
\hypersetup{%
 pdfsubject=Paper,
 pdfkeywords={nuclear physics} 
 unicode = true,
 breaklinks = true,
 colorlinks = true,
 linkcolor = blue,
 menucolor = blue,
 citecolor = blue,
 urlcolor = blue
}

\begin{document}
\title{Determination of Mid-Infrared Refractive Indices of Superconducting Thin Films Using Fourier Transform Infrared Spectroscopy}

\author{Dip Joti Paul}
\email{djpaul@mit.edu}
\author{Tony X. Zhou}
\altaffiliation{Present address: Northrop Grumman Mission Systems, Linthicum, Maryland 21090, USA} 
\author{Karl K. Berggren} 
\affiliation{Research Laboratory of Electronics, Massachusetts Institute of Technology, Cambridge, MA 02139, USA}

\begin{abstract}
In this work, we present a technique to determine the mid-infrared refractive indices of thin superconducting films using Fourier transform infrared spectroscopy (FTIR). In particular, we performed FTIR transmission and reflection measurements on 10-nm-thick NbN and 15-nm-thick MoSi films in the wavelength range of 2.5 to 25 $\mu$m, corresponding to frequencies of 12–120 THz or photon energies of 50–500 meV. To extract the mid-infrared refractive indices of these thin films, we used the Drude-Lorentz oscillator model to represent their dielectric functions and implemented an optimization algorithm to fit the oscillator parameters by minimizing the error between the measured and simulated FTIR spectra. \highlightblue{We performed Monte Carlo simulations in the optimization routine to estimate error ranges in the extracted refractive indices resulting from multiple sources of measurement uncertainty.} To evaluate the consistency of the extracted dielectric functions, we compared the refractive indices extrapolated from these dielectric functions in the UV to near-infrared wavelengths with the values separately measured using spectroscopic ellipsometry. \highlightblue{We validated the applicability of the extracted mid-infrared refractive indices of NbN and MoSi at temperatures below their critical temperatures by comparing them with the Mattis-Bardeen model.} This FTIR-based refractive index measurement approach can be extended to measure the refractive indices of thin films at wavelengths beyond 25 $\mu$m, which will be useful for designing highly efficient photon detectors and photonic devices with enhanced optical absorption in the mid- and far-infrared wavelengths.
\end{abstract}

\maketitle

Mid-infrared spectroscopy has been instrumental in numerous scientific discoveries and commercial applications due to its unique ability to identify chemical bonds. The photon wavelength, ranging from 2.5 $\mu$m to 25 $\mu$m, is known as the mid-infrared spectrum \cite{R21}. Molecules exhibit distinct absorption and emission patterns within this wavelength range due to the excitation of vibrational and rotational modes in their constituent chemical bonds. Hence, this wavelength spectrum is often referred to as the molecular fingerprint region. Consequently, the development of highly efficient mid-infrared photon detectors has received constant attention in recent years \cite{R17,R5,R13,R36}. 

Compared to conventional semiconducting mid-infrared detectors, superconducting nanowire single-photon detectors (SNSPDs) offer significant advantages in photon-starved environments due to their unique ability to resolve single-photon quanta and maintain extremely low dark count rates \cite{R39}. Therefore, efficient mid-infrared SNSPDs will be valuable in a broad range of applications related to astronomy \cite{R8}, dark matter search \cite{R39,R41}, LIDAR \cite{R10}, biological \cite{R9}, and environmental sensing \cite{R10}. However, the detection efficiency of SNSPDs depends on their photon absorption efficiency and internal detection efficiency. As the wavelength extends into the mid-infrared spectrum, the photon absorption efficiency of SNSPDs drops below 5\%, primarily due to the monotonic decrease in the absorption coefficient of the superconducting material with increasing wavelength \cite{R17, R12, R11}. However, the free-space coupling and absorption of mid-infrared photons with planar \si{\mu m^2}-area SNSPD geometries can be improved by incorporating optical cavities \cite{R13}, impedance-matched antennas \cite{R14}, and metamaterials \cite{R15}. Designing such optical elements, however, requires knowledge of the mid-infrared refractive indices of the SNSPD material. 

In this work, we characterized the mid-infrared optical properties of 10-nm-thick NbN and 15-nm-thick MoSi films. NbN is a polycrystalline superconducting material and it has been widely used in SNSPDs and superconducting circuits \cite{R12,R1}. While there have been previous demonstrations of mid-infrared SNSPDs using NbN \cite{R6,R22}, the photon absorption efficiency was not optimized in those works. In addition, amorphous superconducting materials, such as MoSi and WSi, are considered ideal for sensing low-energy infrared photons \cite{R23,R5,R17}. SNSPDs based on MoSi have been demonstrated to detect 5 \si{\mu m} wavelength photons \cite{R23}, and more recently, WSi thin films have been used to detect photons with a wavelength of 29 \si{\mu m} \cite{R5}. These studies also lack optimization of photon absorption efficiency in SNSPDs, as there have been no reports to date on the mid-infrared optical properties of MoSi and WSi. Therefore, having knowledge of the mid-infrared refractive indices of these superconducting thin films will enable researchers to perform simulations and fabricate optical elements to improve photon absorption efficiency of SNSPDs in the mid-infrared spectrum.

The conventional approach to determining the complex refractive indices of thin-film materials is to use spectroscopic ellipsometry tools. However, commercially available ellipsometry tools typically operate in the wavelength range of visible to near-infrared. The complex refractive indices of NbN and MoSi in the wavelength range from visible to 5 $\mu$m have been reported previously using ellipsometry \cite{R11,R12,R27}. However, finding a commercial ellipsometry tool that can operate across a broad range of wavelengths, from visible to far-infrared, is challenging due to the need to adjust the optical setup of the instrument, including the source, lenses, and detectors, as the wavelength increases. FTIR spectroscopy, on the other hand, is well-suited for infrared spectroscopy and is widely available, with capabilities extending to 50 $\mu$m or beyond. FTIR-based refractive index measurements have been reported for various materials, including organic polymers, ceramics, and semiconductors, across wavelengths up to 100 $\mu$m \cite{R29,R42,R26}. \highlightblue{While this work focuses on characterizing superconducting thin films in the 2.5 to 25 $\mu$m wavelength range, the methodology for extracting complex refractive indices from reflection and transmission spectra is broadly applicable to both bulk crystals and thin films at any wavelength \cite{R44}.} Hence, this FTIR-based approach can be valuable for a wide range of photonic devices and applications by reducing dependence on specialized ellipsometry tools for refractive index measurements beyond the typical wavelength range of commercially available instruments.

In this work, we developed an optimization algorithm to extract the complex refractive indices of thin-film NbN and MoSi from their FTIR transmission and reflection spectra. \highlightblue{We incorporated measurement uncertainties in the FTIR spectra and thin-film thickness into the optimization routine using Monte Carlo analysis and reduced chi-squared error function, and calculated the resulting uncertainty bounds in the extracted refractive indices.} We validated the consistency of the extracted Drude-Lorentz parameters using spectroscopic ellipsometry and DC resistivity measurements. Finally, \highlightblue{we validated the applicability of the extracted room-temperature mid-infrared refractive indices of NbN and MoSi at temperatures below their critical temperatures by comparing them with simulated values obtained from the Mattis-Bardeen model \cite{R4}.}

\begin{figure}[!b]
\vspace{-0.35cm}
\includegraphics[width=\linewidth]{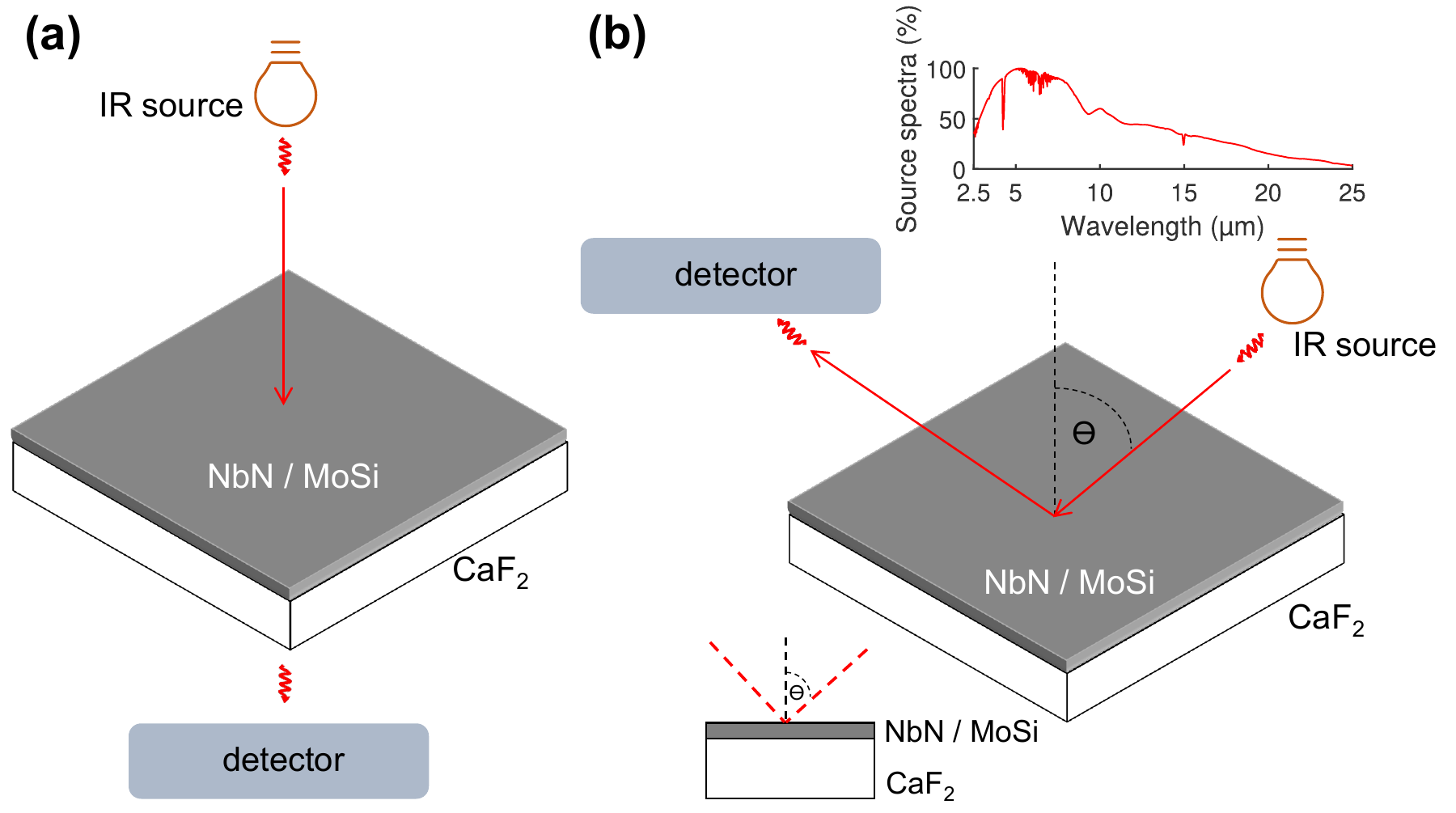}
\vspace{-0.5cm}
\caption{\footnotesize A schematic representation of FTIR spectroscopy performed on the deposited NbN and MoSi films on (100)-oriented \si{CaF_2} substrate is shown. (a) Normal incidence FTIR transmission measurement and (b) variable incidence angle FTIR reflection measurements were performed. The incidence angle, $\theta$, was varied from 12° to 60° in the reflection measurements. FTIR source spectra are shown in the inset, with KBr used as the beam splitter and DTGS as the detector in the FTIR instrument.}
\label{fig:Fig1}
\end{figure}

In this work, we deposited NbN and MoSi films on (100)-oriented \si{CaF_2} substrates at room temperature using an AJA Orion magnetron sputtering system. \si{CaF_2} was chosen as the substrate due to its optical transparency in the near- and mid-infrared wavelengths \cite{R2}. The infrared transparency of the substrate allowed us to measure the transmitted spectrum of the deposited thin films, which would not be possible with an opaque substrate \highlightblue{such as oxidized silicon substrate}. Undoped silicon is also a viable substrate, as it is optically transparent in the near- and mid-infrared wavelengths. We used the same optimized deposition conditions for NbN and MoSi thin films as reported in our previous works \cite{R1,R25}. \highlightblue{A detailed description of the deposition conditions and the effect of the substrate on the deposited NbN and MoSi films is provided in supplementary section A.}



The thicknesses of the deposited NbN and MoSi films on \si{CaF_2} substrates were measured using X-ray reflectivity (XRR) and determined to be 9.93 nm and 14.98 nm, respectively. While FTIR can measure films of varying thickness, we characterized these thickness values for the NbN and MoSi films, as they fall within the typical range of superconducting film thicknesses used in SNSPD fabrication \cite{R6,R25}. The sheet resistances of the deposited NbN and MoSi films were 240.68 \si{\ohm}/sq and 110.51 \si{\ohm}/sq, respectively. The samples were stored in an inert atmosphere box to minimize oxidation of NbN and MoSi. 

We performed FTIR reflection and transmission measurements on the samples using a Thermo Scientific Nicolet iS50. A schematic of the measurement setup is shown in Figure \ref{fig:Fig1}. In the reflection measurements, the reflected spectra from the sample were normalized by the reflected spectra of a gold mirror, while in the transmission measurements, source spectra were used to normalize the sample’s transmitted spectra. \highlightblue{We used a Thorlabs 25.4 mm diameter unprotected gold mirror (PF10-03-M03) in this work, which, as specified in the datasheet, has an absolute reflectance above 97\% over the 800 nm to 20 $\mu$m wavelength range. This specification was verified using a PerkinElmer UV-VIS-NIR spectrophotometer, with details provided in supplementary section B.} All measurements were performed with an unpolarized collimated beam, and we conducted averaging of 64 scans per measurement with automatic atmospheric suppression enabled. To improve the accuracy of the dielectric fitting model, we measured several spectra of the sample, including transmittance at normal incidence and reflectance at incidence angles ranging from 12° to 60°. We used the Harrick specular reflection accessory for the 12° incidence angle and the Pike VeeMAX III for incidence angles from 30° to 60°. \highlightblue{In addition, we performed several measurements to quantify the statistical 1$\sigma$ standard uncertainty in the measured FTIR spectra, as discussed in supplementary section B.}

\begin{figure}[!t]
\vspace{-0.5cm}
\includegraphics[width=\linewidth]{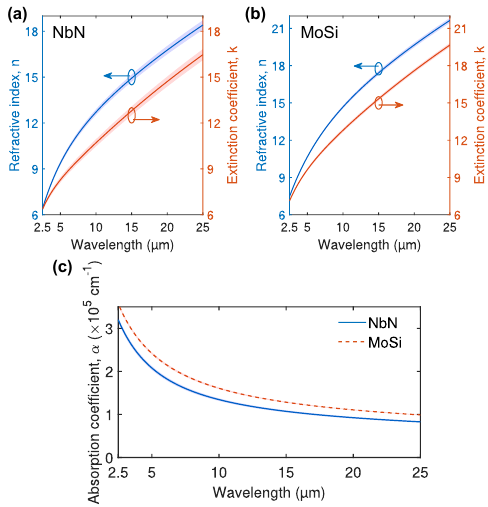}
\vspace{-0.5cm}
\caption{\footnotesize Extracted mid-infrared refractive index and extinction coefficient of (a) 10-nm-thick NbN and (b) 15-nm-thick MoSi \protect\highlightblue{(solid line: mean values; shaded region: uncertainty bounds)}. (c) The absorption coefficient of the materials is calculated using the equation $\alpha = 4\pi k / \lambda$, where $\lambda$ is the wavelength and $k$ is the extinction coefficient. \protect\highlightblue{The shaded region represents the uncertainty bounds arising from the uncertainty in the extracted extinction coefficient.}}
\label{fig:Fig2}
\vspace{-0.5cm}
\end{figure} 

To extract the complex refractive indices of NbN and MoSi from the measured FTIR spectra, we developed an optimization algorithm based on the transfer matrix method \cite{R28}. \highlightblue{To validate that the transfer matrix simulation model applies to our FTIR measurement setup, we performed a spectral energy conservation analysis, as detailed in supplementary section D.} The Drude-Lorentz oscillator model was used to describe the dielectric functions of NbN and MoSi films, as their room-temperature electrical conductivity lies between good metals and semiconductors \cite{R12}. The model is given by the following equation: \vspace{-0.15cm}
\begin{equation*}
\epsilon(\omega) = \epsilon_{\infty} - \frac{\omega_p^2}{\omega^2 + i \omega \Gamma_{D}} + \sum_{k=1}^N \frac{s_k^2}{\omega_k^2 - \omega^2 - i \omega \gamma_k}
\vspace{-0.1cm}
\end{equation*} 
where $\omega$ is the spectral frequency, and $\epsilon_{\infty}$ is the high-frequency dielectric constant of the material. $\omega_p$ and $\Gamma_D$ are the fitting parameters of the Drude model, with $\omega_p$ representing the unscreened plasma energy and $\Gamma_D$ as the inverse of the free-electron relaxation time. The number of Lorentz oscillators is defined by N, where $s_k$, $\omega_k$, and $\gamma_k$ are the strength, resonant frequency, and damping coefficient of the \textit{k}-th Lorentz oscillator, respectively. In our calculation, the frequency ($\omega$) and the Drude-Lorentz oscillator parameters were taken in units of electron-volt. 

Following the calculation of the Drude-Lorentz dielectric function, the complex refractive index of the material was determined by $\overline{n(\omega)}= n(\omega)+ik(\omega)= \sqrt{\epsilon(\omega)}$ where $n(\omega)$ and $k(\omega)$ are the refractive index and extinction coefficient, respectively. Our dielectric fitting algorithm employed a nonlinear optimization function to determine the optimal values of the Drude-Lorentz parameters from an initial guess by iteratively minimizing the error between the measured FTIR spectra and the simulated spectra computed with the fitted dielectric function over the wavelength range of interest. \highlightblue{We implemented Monte Carlo approach to propagate the uncertainties in several measured quantities, such as FTIR spectra, incidence angle, and film thicknesses, to the best-fit Drude-Lorentz parameters of the NbN and MoSi films. A flowchart of the optimization routine with detailed information is provided in supplementary Figure S3}.

\begin{figure}[!t]
\vspace{-0.5cm}
\includegraphics[width=\linewidth]{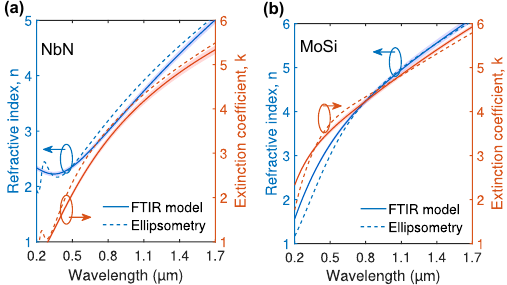}
\vspace{-0.5cm}
\caption{\footnotesize Refractive indices of (a) 10-nm-thick NbN and (b) 15-nm-thick MoSi films \protect\highlightblue{on \si{CaF_2}} in the UV to near-infrared. Solid lines show extrapolated values from FTIR-extracted Drude–Lorentz parameters, \protect\highlightblue{with shaded regions indicating uncertainty bounds in the refractive indices due to uncertainties in the extracted parameters}. Dotted lines represent ellipsometry measurements.}
\label{fig:Fig3}
\vspace{-0.5cm}
\end{figure}

To determine the refractive indices of the deposited thin films of NbN and MoSi from the FTIR spectra, the dielectric fitting algorithm requires the dielectric function of the substrate. Although there are reported values of the refractive index of \si{CaF_2} in the literature \cite{R43,R2}, we chose to determine it using our FTIR-based refractive index measurement method. This measurement allowed us to compare the results from our refractive index measurement method with the literature. Hence, we measured the FTIR transmission and reflection spectra of \highlightblue{a double-sided polished, 500 $\mu$m-thick} \si{CaF_2} substrate that we used to deposit NbN and MoSi thin films. We then applied our dielectric fitting algorithm to extract the complex refractive indices of \si{CaF_2} from the measured FTIR spectra. The extracted refractive index values are provided in the \highlightblue{supplementary Figure S4(b)}, which showed good agreement with \cite{R43,R2}. \highlightblue{Details of the dielectric fitting of \si{CaF_2} are provided in supplementary section C.} 

\begin{table*}[htbp]
    \centering
    \begin{tabularx}{\textwidth}{*{12}{X}}
        \toprule
        Sample &  & $\epsilon_{\infty}$ & $\omega_p$ & $\Gamma_{D}$ & $N$ & $s_1$ & $\omega_1$ & $\gamma_1$ & $s_2$ & $\omega_2$ & $\gamma_2$\\
        \midrule 
        NbN & \protect\highlightblue{mean} & 6.91 & 6.94 & 1.66 & 1 & 18.14 & 1.98 & 21.41 \\
        & \protect\highlightblue{$\pm$ std} & 0.195 & 0.096 & 0.047 &  & 0.317 & 0.046 & 0.757 \\
        MoSi & \protect\highlightblue{mean} & 1.15 & 15.01 & 5.53 & 2 & 20.61 & 5.01 & 15.5 & 3.24 & 0.36 & 1.02 \\
        & \protect\highlightblue{$\pm$ std} & 0.02 & 0.082 & 0.059 &  & 0.353 & 0.198 & 0.29 & 0.03 & 0.018 & 0.026 \\
        \bottomrule
    \end{tabularx}
    \vspace{-0.25cm}
    \caption{Extracted Drude-Lorentz oscillator parameters for 10-nm-thick NbN and 15-nm-thick MoSi films, \protect\highlightblue{obtained from Monte Carlo simulations} based on the best fit to the measured FTIR data. The parameters are in units of electron-volts (eV).}
    \label{tab:T1}
    \vspace{-0.5cm}
\end{table*}

Next, we measured the FTIR spectra of 10-nm-thick NbN sample deposited on a \si{CaF_2} substrate and determined the refractive indices of NbN between 2.5 $\mu$m and 25 $\mu$m wavelength, as shown in Figure \ref{fig:Fig2}(a). Similarly, we performed FTIR measurements on 15-nm-thick MoSi deposited on a \si{CaF_2} substrate and then extracted the refractive indices of MoSi, as shown in Figure \ref{fig:Fig2}(b). \highlightblue{We used the reduced chi-squared error function in our optimization algorithm to incorporate the measurement uncertainties. The final reduced chi-squared error values for the optimized parameters were 1.139 for NbN and 1.206 for MoSi. The reduced chi-squared errors being slightly greater than one in both cases are likely due to underestimated measurement uncertainties in the incidence angle and substrate thickness, as these uncertainties were not accurately characterized. The simulated spectra, generated using the extracted refractive indices and their associated uncertainty bounds, were compared with the FTIR measurements, as shown in supplementary Figure S4.} In addition, the \highlightblue{mean values and standard deviations of the} extracted Drude-Lorentz parameters for NbN and MoSi are given in Table \ref{tab:T1}. 

We examined the wavelength-dependent mid-infrared optical absorption in these thin films by calculating their absorption coefficients, as shown in Figure \ref{fig:Fig2}(c). As depicted in Figure \ref{fig:Fig2}(c), the mid-infrared absorption coefficients of NbN and MoSi decrease with increasing wavelength, indicating that the photon absorption efficiency of NbN- and MoSi-based SNSPDs will monotonically decrease as the photon wavelength extends into the mid-infrared spectrum.

To evaluate whether the extracted Drude-Lorentz parameters for NbN and MoSi are not overfitted to the specified mid-infrared wavelength range but are applicable across a broader spectrum, we compared the refractive indices extrapolated from these parameters in the wavelength range of 0.2 to 1.7 $\mu$m with values independently measured using the Semilab SE-2000 spectroscopic ellipsometry. The ellipsometry tool relies on reflection spectroscopy to measure changes in the polarization of reflected light at various incidence angles and wavelengths. Following the measurements, we determined the unknown refractive index of the material using an integrated analysis software by performing a best-fit optimization with various dielectric functions available in the software, such as Drude, Lorentz, Gauss, Sellmeier, and Cauchy functions. The Drude-Lorentz dielectric model was identified as the optimal model to represent the optical properties of NbN and MoSi. These measured refractive indices were then compared with the extrapolated refractive indices from our FTIR measurements, as shown in Figures \ref{fig:Fig3}(a) and \ref{fig:Fig3}(b). \highlightblue{Since the ellipsometry tool is based on reflectometry, we encountered challenges in obtaining accurate models for NbN and MoSi films deposited on the transparent \si{CaF_2} substrate. The R-squared (\si{R^2}) error for the ellipsometric fitting was 0.936 for NbN and 0.939 for MoSi on the \si{CaF_2} substrate.} The refractive indices extrapolated from the FTIR-extracted Drude-Lorentz parameters matched reasonably well with the ellipsometric data, suggesting that the parameters perform well across the UV to mid-infrared.

Next, we calculated the plasma wavelength and DC sheet resistance of the NbN and MoSi films using the FTIR-extracted Drude parameters to assess whether these parameters are physically meaningful rather than merely fitting parameters. The plasma wavelengths of the NbN and MoSi films can be estimated from their Drude parameter $\omega_p$ using $\lambda_{p} [nm] = 1239.8/\omega_p [eV]$. We calculated the plasma wavelength to be \highlightblue{82.6 $\pm$ 0.37} nm for 15-nm-thick MoSi and \highlightblue{178.8 $\pm$ 2.02} nm for 10-nm-thick NbN. The plasma wavelength of 11.7-nm-thick NbN was reported as 139.1 nm in the literature \cite{R12}, which is consistent with our calculated value.

In addition, the DC resistivity of the films can be calculated using $\rho_n = (\Gamma_{D} \hbar)/(\omega_p^2 \epsilon_0 e)$. Here, $\epsilon_0$ is the free-space permittivity, $e$ is the electron charge, and $\hbar$ is the reduced Planck constant. \highlightblue{The derivation of the equation is provided in supplementary section E.} The Drude parameter values, $\omega_p$ and $\Gamma_D$, for NbN and MoSi, were taken from Table \ref{tab:T1}. We calculated the DC resistivity ($\rho_n$) of the 10-nm-thick NbN and 15-nm-thick MoSi films and then determined their corresponding sheet resistances using $R_{sheet} = \rho_n/\text{thickness}$. This resulted in sheet resistances of \highlightblue{258.47 $\pm$ 10.27} \si{\ohm}/sq and \highlightblue{121.98 $\pm$ 1.86} \si{\ohm}/sq for the NbN and MoSi films, respectively, which are close to the measured room-temperature values of 240.68 \si{\ohm}/sq and 110.51 \si{\ohm}/sq. Hence, we infer that the extracted Drude-Lorentz parameters for thin-film NbN and MoSi demonstrate physical consistency and effectively describe their optical behavior across the UV to mid-infrared.


In this work, we determined the mid-infrared refractive indices of NbN and MoSi using room-temperature FTIR spectroscopy; however, our primary interest lies in the optical properties of these materials in their superconducting state for performing optical simulations of SNSPDs and other superconducting photonic devices. The prevailing view in the literature is that the optical constants of a superconductor below $T_\mathit{c}$ do not deviate substantially from their room-temperature values when the photon energy is well above the superconducting gap energy \cite{R11, R12}. The superconducting gap energy of thin-film NbN and MoSi can be estimated from their critical temperatures using the relation $3.52 \times k_B T_\mathit{c}$ \cite{R3}, where $k_B$ is the Boltzmann constant and $T_\mathit{c}$ is the critical temperature of the superconducting thin film. \highlightblue{The critical temperatures of the deposited NbN and MoSi thin films were 9.67 K and 5.23 K, corresponding to the superconducting gap energies of 2.93 meV and 1.59 meV, respectively. These energies correspond to photon wavelengths of about 400 $\mu$m and 800 $\mu$m, which are much longer than the wavelengths considered in this work.} Therefore, the refractive indices extracted in the wavelength range of 2.5 to 25 $\mu$m using FTIR measurements should closely approximate the refractive index at temperatures below $T_\mathit{c}$ and hence be suitable for use in the design of superconducting photonic devices. \highlightblue{Further discussion is provided in supplementary section F, where we calculated the refractive indices of thin-film NbN and MoSi at 4 K using the Mattis-Bardeen model \cite{R4} and compared them with the FTIR-extracted room-temperature refractive indices.}

In summary, we determined the mid-infrared refractive indices of 10-nm-thick NbN and 15-nm-thick MoSi using room-temperature FTIR spectroscopy. The accuracy of the extracted optical properties was validated with spectroscopic ellipsometry and DC sheet resistance measurements. While optimization of the stoichiometry and thickness of NbN and MoSi films may be required to enhance the mid-infrared photon detection sensitivity of SNSPDs, the results presented here provide a useful foundation for characterizing and understanding the optical properties of these films. Furthermore, this approach for determining refractive indices using FTIR spectroscopy will be effective for extracting the optical properties of thin films of various materials in the mid- to far-infrared wavelengths, where obtaining a specialized ellipsometry tool for that wavelength range may be challenging.

This work was carried out in part through the use of MIT.nano's facilities, and the authors would like to thank Tim McClure of the MIT Materials Research Laboratory (MRL) for the technical support related to FTIR. The authors also thank Francesca Incalza, Matthew Yeung, and Phillip D. Keathley for their feedback during the preparation of the manuscript. The initial stage of this research was sponsored by the US Army Research Office (ARO) and was accomplished under Cooperative Agreement No. W911NF-21-2-0041. The later stages of this research were supported by MIT Lincoln Laboratory under the SNSPD-SFQ Line program and by the National Science Foundation under Grant No. EEC-1941583.

Source codes used to extract the refractive indices of NbN and MoSi from FTIR measurements are available on GitHub (https://github.com/DipJotiPaul/RefractEx). 
     
%

\end{document}


\title{Supplementary Information \\ Determination of Mid-Infrared Refractive Indices of Superconducting Thin Films Using Fourier Transform Infrared Spectroscopy}
\author{Dip Joti Paul}
\author{Tony X. Zhou}
\author{Karl K. Berggren} 
\affiliation{Research Laboratory of Electronics, Massachusetts Institute of Technology, Cambridge, MA 02139, USA}
\maketitle

\section*{Methods}
\vspace{-0.6cm}
\subsection{Deposition of NbN and MoSi Thin Films}
\vspace{-0.5cm}
\highlightblue{In this work, we deposited NbN and MoSi films on \si{CaF_2} and oxidized silicon substrates using an AJA Orion magnetron sputtering system in a cryopumped chamber at base pressure of 5$\times 10^{-9}$ Torr. \si{CaF_2} was chosen for its mid-infrared transparency, while oxidized silicon served as a reference substrate to characterize the deposition consistency of the films, since the sputtering conditions for these materials have been previously optimized on oxidized silicon substrate in our chamber for fabricating high-performance SNSPDs. For the NbN film deposition, the sputtering parameters were: \si{N_2} flow of 6 sccm, Ar flow of 26.5 sccm, 17 W RF bias to the substrate holder, and 160 W DC power to the Nb target. Deposition was performed for 245 seconds at room temperature, and the thickness and sheet resistance of the resulting films on both substrates were subsequently measured, as given in Table \ref{tab:S1}. The MoSi films were deposited by co-sputtering from Mo and Si targets at an Ar flow of 26.5 sccm, with 120 W DC power applied to the Mo target and 120 W RF power to the Si target. Deposition was performed for 400 seconds at room temperature, and the thickness and sheet resistance of the resulting films are given in Table \ref{tab:S1}.}

\begin{figure}[!t]
\vspace{-0.2cm}
\includegraphics[width=0.85\linewidth]{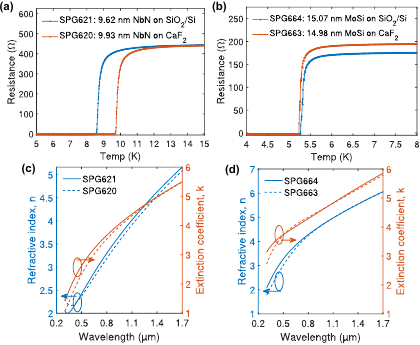}
\vspace{-0.55cm}
\caption{\footnotesize \protect\highlightblue{Measurement of the superconducting transition temperature (\si{T_{\mathit{c}}}) of the deposited (a) 10-nm-thick NbN and (b) 15-nm-thick MoSi films on \si{CaF_2} and oxidized silicon substrates. The refractive indices of (c) 10-nm-thick NbN and (d) 15-nm-thick MoSi, obtained from ellipsometry measurements, are also shown. Solid lines represent films on oxidized silicon substrate, while dotted lines represent films on \si{CaF_2}.}} \vspace{-0.25cm}
\label{fig:FigS3}
\end{figure}

\highlightblue{Thickness of the deposited films was determined from the XRR curves using Rigaku’s GlobalFit software. The root mean square (RMS) film roughness of the NbN films on the \si{CaF_2} and oxidized silicon substrates was 0.6 nm and 0.27 nm, respectively, and the deposition rate of NbN on the \si{CaF_2} substrate was found to be slightly higher than on the oxidized silicon substrate. The superconducting transition temperature of the deposited NbN on the \si{CaF_2} and oxidized silicon substrates was measured, as shown in Figure \ref{fig:FigS3}(a).}

\highlightblue{The RMS film roughness of the MoSi films on the \si{CaF_2} and oxidized silicon substrates was 0.65 nm and 0.58 nm, respectively, and the deposition rate of MoSi on the \si{CaF_2} substrate was found to be nearly the same as on the oxidized silicon substrate. The superconducting transition temperature of the deposited MoSi on the \si{CaF_2} and oxidized silicon substrates was measured, as shown in Figure \ref{fig:FigS3}(b).}

\begin{table*}[htbp]
    \centering
    \begin{tabularx}{0.8\textwidth}{X X X X X X X}
        \toprule
        Sample & Film & Substrate & Deposition Time & Thickness & $R_{\text{sheet}}$ ($\Omega/\square$) & $T_c$ (K)\\
        \midrule
        SPG621 & NbN & SiO$_2$/Si & 245 s & 9.62 nm & 246.44 & 8.52 \\
        SPG620 & NbN & CaF$_2$ & 245 s & 9.93 nm & 240.68 & 9.67 \\
        SPG664 & MoSi & SiO$_2$/Si & 400 s & 15.07 nm & 109.91 & 5.27 \\
        SPG663 & MoSi & CaF$_2$ & 400 s & 14.98 nm & 110.51 & 5.23 \\
        \bottomrule
    \end{tabularx}
    \vspace{-0.05cm}
    \caption{\protect\highlightblue{Properties of the NbN and MoSi thin films deposited on \si{CaF_2} and oxidized silicon substrates, as measured in this work.}}
    \label{tab:S1}
    \vspace{-0.25cm}
\end{table*}

\highlightblue{We performed room-temperature ellipsometry measurements of the deposited NbN and MoSi films on the \si{CaF_2} and oxidized silicon substrates, as shown in Figures \ref{fig:FigS3}(c) and \ref{fig:FigS3}(d). Since the ellipsometry tool is based on reflectometry, we encountered challenges in obtaining accurate models for NbN and MoSi films deposited on the transparent \si{CaF_2} substrate. The R-squared (\si{R^2}) error for the ellipsometric fitting was 0.936 and 0.989 for NbN on \si{CaF_2} and oxidized silicon, respectively, while for MoSi on \si{CaF_2} and oxidized silicon, the values were 0.939 and 0.975, respectively. Overall, we find that for amorphous MoSi thin film deposition, both \si{CaF_2} and oxidized silicon substrates yielded similar film properties. In contrast, NbN deposited on \si{CaF_2} showed a higher deposition rate and critical temperature (\si{T_{\mathit{c}}}) compared to the film on oxidized silicon, likely due to the crystalline structure of \si{CaF_2}, which facilitates higher-quality NbN growth.}

\begin{figure}[!b]
\vspace{-0.5cm}
\includegraphics[width=0.85\linewidth]{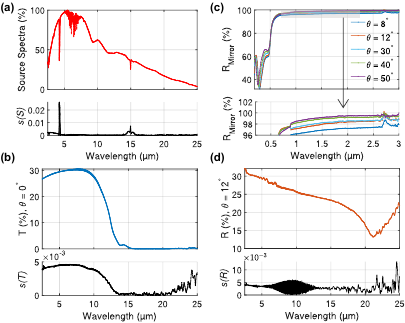}
\vspace{-0.5cm}
\caption{\footnotesize \protect\highlightblue{Measurement uncertainty of the FTIR spectra. We performed six repeated measurements for each of two sample placement configurations and plotted the mean values in the top figure panel, with the statistical 1$\sigma$ standard uncertainty shown in the bottom figure panel for (a) FTIR source spectra measured without any sample between the source and detector, (b) transmittance spectra of 10-nm-thick NbN on \si{CaF_2}, and (d) reflectance spectra of 10-nm-thick NbN on \si{CaF_2} at 12$^\circ$ incidence angle. (c) Absolute reflectance of the reference gold mirror measured at different incidence angles ($\theta$) using a UV-VIS-NIR spectrophotometer. The bottom figure panel shows a zoomed-in view.} \vspace{-0.5cm}}
\label{fig:FigS4}
\end{figure}

\vspace{-0.9cm}
\subsection{Uncertainty in FTIR Measurements}
\vspace{-0.5cm}
\highlightblue{Uncertainty in FTIR measurements can arise from instrumental, environmental, and sample-related factors. For example, fluctuations in the intensity of the FTIR light source, thermal or electronic noise in the detector, and surface roughness or thickness nonuniformity of the sample can all affect the measured spectra. To reduce the instrumental and environmental variability, we selected the option to average 64 scans per measurement in the FTIR software for each measurement in this work. We then performed six repeated measurements without changing the measurement setup or sample placement, which allowed us to quantify the uncertainty arising from instrumental and environmental variations. To account for sample-related variability, we performed six additional measurements after slightly repositioning the sample from its original placement.}

\highlightblue{To measure the transmittance of the sample, we performed the ratio of the sample’s transmittance spectra to the source spectra measured without any sample in the holder. Figure \ref{fig:FigS4}(a) shows the mean and 1$\sigma$ standard uncertainty of six repeated measurements of the source spectra. The 1$\sigma$ standard uncertainty quantifies the variability introduced by instrumental and environmental factors; however, the uncertainty remained below 0.5\% for the transmittance of the 10-nm-thick NbN sample, as shown in Figure \ref{fig:FigS4}(b). We performed six measurements for each of the two slightly shifted sample positions, thereby incorporating sample-related variability into this uncertainty metric. Similar transmittance measurements were performed on the MoSi sample, and a comparable 1$\sigma$ uncertainty was observed.}

\highlightblue{To normalize the reflectance spectra of the sample in the FTIR measurements, we used a Thorlabs 25.4 mm diameter unprotected gold mirror (PF10-03-M03). According to the Thorlabs datasheet, the reflectance of unpolarized light from this mirror is approximately 98.75\% at 45$^\circ$ incidence angle over the 800 nm to 20 $\mu$m wavelength range. We also measured the absolute reflectance of the gold mirror at several incidence angles using a PerkinElmer Lambda 1050 UV-VIS-NIR spectrophotometer in the wavelength range of 0.2 to 3 $\mu$m, as shown in Figure \ref{fig:FigS4}(c). This measurement verified the approximate 98.75\% absolute reflectance specification of the reference gold mirror provided in the datasheet. To account for the non-unity absolute reflectance of the reference mirror, we multiplied the normalized FTIR reflectance spectra of the sample by the average value of our measured absolute reflectance of the gold mirror over the 2.5 to 3 $\mu$m wavelength range for each corresponding incidence angle. In addition, the mirror was handled with care to avoid scratching its reflective surface.}

\highlightblue{Figure \ref{fig:FigS4}(d) shows the reflectance uncertainty of the NbN sample at 12$^\circ$ incidence, normalized using the reflected spectra of the gold mirror. Sample positional variability was considered by performing six repeated measurements at each sample position to quantify the 1$\sigma$ standard uncertainty, which indicates that the worst-case uncertainty in our reflectance measurements can be as high as 1\%.}

\begin{figure}[!b]
\vspace{-0.5cm}
\includegraphics[width=0.9\linewidth]{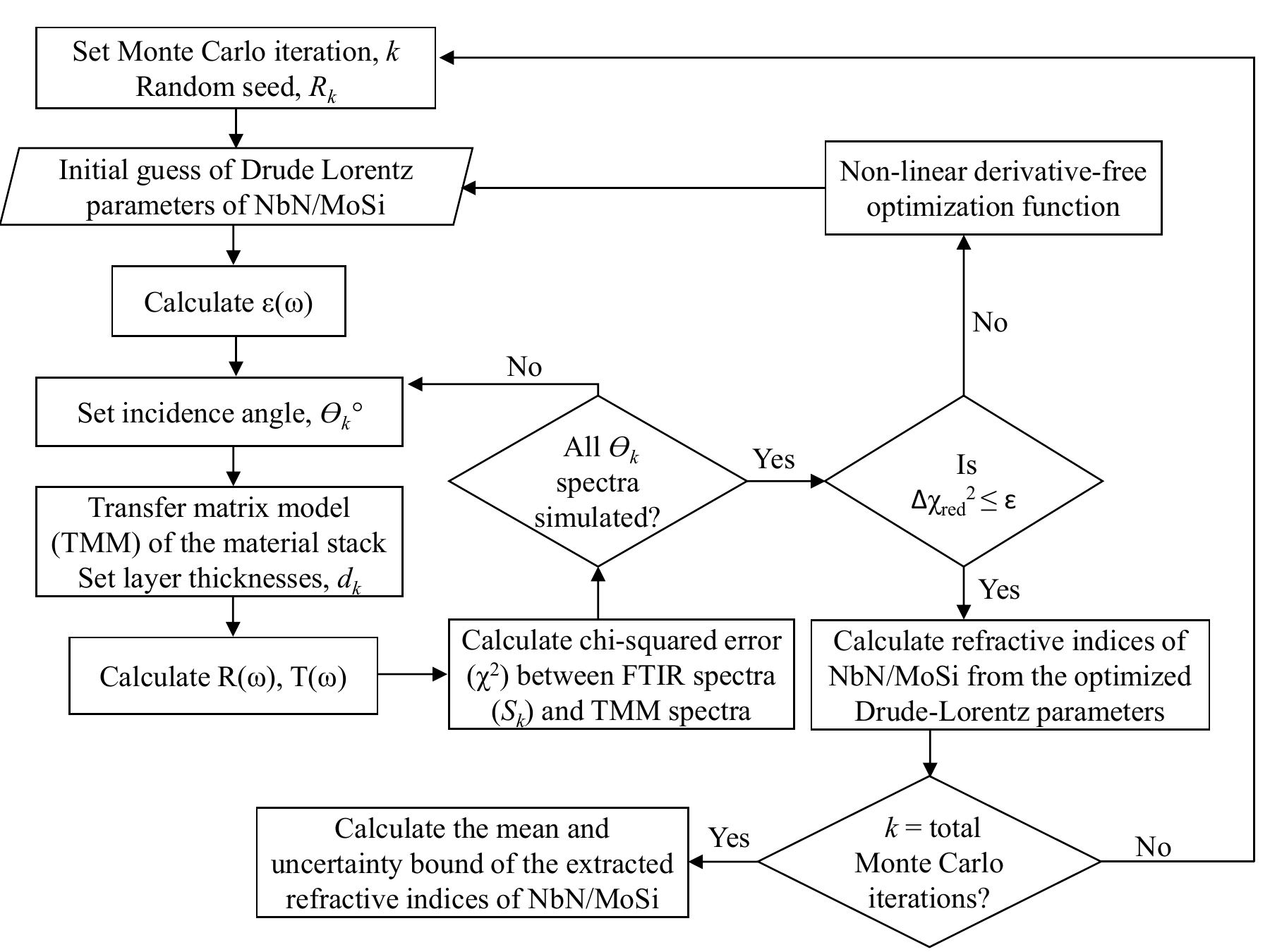}
\vspace{-0.75cm}
\caption{\footnotesize A flowchart of the dielectric fitting optimization algorithm is shown, which was employed to extract the refractive indices of the NbN and MoSi thin films from their measured FTIR spectra. \vspace{-0.2cm}}
\label{fig:FigS1}
\end{figure}

\vspace{-0.9cm}
\subsection{Extraction of Refractive Index from FTIR Spectra}
\vspace{-0.5cm}
To extract the refractive indices of NbN and MoSi from the measured FTIR spectra, we developed a dielectric constant fitting program based on the transfer-matrix method \cite{R28}. The program can simulate the transmission and reflection spectra of a sample at specified incidence angles, polarization angles, and wavelengths, given that the thickness and refractive indices of the constituent materials of the sample are known at those wavelengths. To extract the refractive index of a material from its FTIR spectra, we added an optimization routine to the program, as shown in Figure \ref{fig:FigS1}. The input data for the program include the FTIR transmission and reflection spectra of the sample, the incidence angle values of the corresponding spectra, and the thickness of the different material layers in the sample. All our FTIR measurements were performed with an unpolarized beam. Additionally, the refractive indices of the known materials in the sample stack were fed into the program, and we assumed a suitable dielectric constant model for the unknown material. In this work, we employed the Drude-Lorentz dielectric model to describe the optical properties of NbN and MoSi. Since the Drude-Lorentz dielectric model inherently ensures Kramers-Kronig (KK) consistency between the real and imaginary parts of the optical constants \cite{R37,R38}, a separate KK analysis was not performed in the dielectric fitting optimization algorithm.

\highlightblue{However, uncertainties exist in several measured quantities, including the FTIR transmission and reflection spectra, incidence angle of light, and the thicknesses of the constituent layers in the material stack. To propagate these uncertainties to the best-fit Drude-Lorentz parameters of the NbN and MoSi films, we used Monte Carlo approach \cite{R26}. We performed one hundred Monte Carlo iterations to evaluate the variation in the extracted refractive indices of NbN and MoSi due to measurement uncertainties, as shown in Figure \ref{fig:FigS1}. One hundred iterations were used in this work to complete the simulation within a reasonable time-frame, as each iteration took approximately 5 minutes to run on a standard user laptop.}

\highlightblue{In each iteration, we randomly sampled a realization of the measurement values, such as the FTIR spectra, layer thicknesses, and incidence angle, from their respective uncertainty distributions. The uncertainty distribution of the FTIR transmission and reflection spectra was quantified, as shown in Figure \ref{fig:FigS4}. The XRR thickness uncertainty was set equal to the RMS film roughness, measured as 0.6 nm for NbN and 0.65 nm for MoSi on the \si{CaF_2} substrate, while the FTIR reflection incidence angle uncertainty was assumed to be $\pm 1^\circ$. We then applied the dielectric fitting optimization routine to each realization, yielding the best-fit Drude-Lorentz parameters of the unknown material for that particular set of measurement data. By repeating this process across many iterations, we obtained the refractive indices of NbN and MoSi with error ranges that reflect the effects of measurement uncertainties.}
\begin{figure}[!t]
\vspace{-0.5cm}
\includegraphics[width=\linewidth]{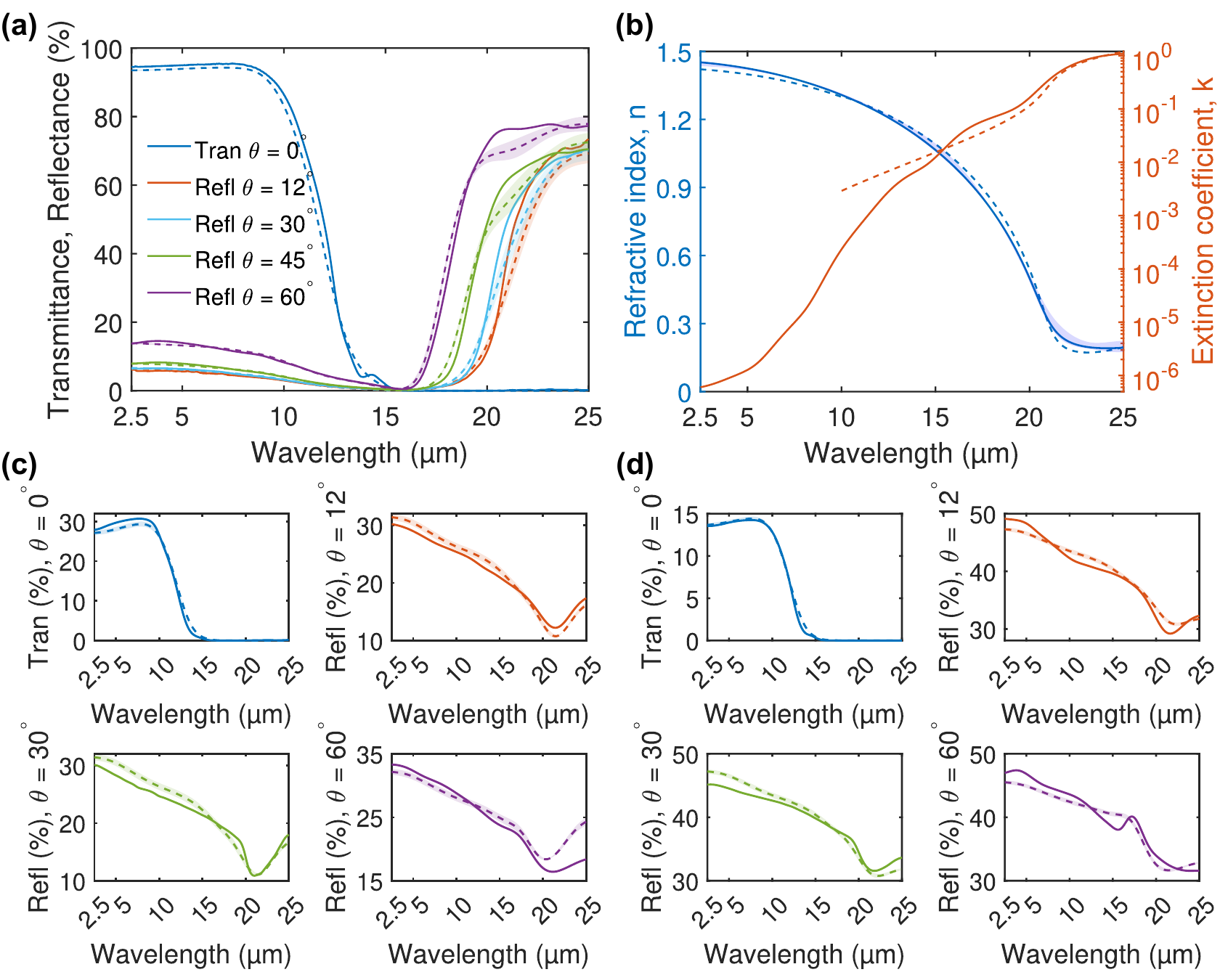}
\vspace{-1.5cm}
\caption{\footnotesize (a) Transmittance and reflectance spectra of bare \si{CaF_2} measured by FTIR (solid line) and simulated using the extracted refractive indices (dashed line: \protect\highlightblue{mean values, shaded region: uncertainty bounds}). (b) Refractive indices of \si{CaF_2} extracted from our dielectric fitting (solid line: \protect\highlightblue{mean values, shaded region: uncertainty bounds}) compared with literature values \cite{R43,R2} (dashed line). (c) FTIR transmittance and reflectance spectra (solid line) of 10-nm-thick NbN on \si{CaF_2}, along with corresponding simulations using the extracted refractive indices of NbN and \si{CaF_2} (dashed line: \protect\highlightblue{mean values, shaded region: uncertainty bounds}). (d) FTIR spectra (solid line) and corresponding simulated spectra (dashed line: \protect\highlightblue{mean values, shaded region: uncertainty bounds}) of 15-nm-thick MoSi on \si{CaF_2}. \vspace{-0.5cm}}
\label{fig:FigS2}
\end{figure}

\highlightblue{The dielectric fitting optimization routine uses a nonlinear derivative free method, the Nelder-Mead algorithm, to determine the final set of Drude-Lorentz parameters from an initial guess by minimizing the reduced chi-squared error ($\chi^2_{red}$) over the wavelength range of interest. The reduced chi-squared function is given by 
\begin{equation*}
\chi^2_{red} = \frac{1}{M \times N - p} \sum_{j=1}^M \sum_{\lambda=2.5\,\mu\text{m}}^{25\,\mu\text{m}} \left( \frac{X_{\text{meas},j}(\lambda) - X_{\text{calc},j}(\lambda)}{\sigma_j(\lambda)} \right)^2
\end{equation*}
where $M$ is the total number of FTIR spectra measured for the sample, $N$ is the number of wavelength points in each spectrum, $p$ is the number of fitted parameters, $\sigma_{j}(\lambda)$ is the measured uncertainty at wavelength $\lambda$ for spectrum $j$, and $X_{(meas,j)}(\lambda)$ and $X_{(calc,j)}(\lambda)$ are the transmittance or reflectance spectra from the FTIR measurement and calculation, respectively. We used one transmission spectrum and several reflection spectra of the sample, measured at various incidence angles, to perform the dielectric fitting optimization, as shown in Figure \ref{fig:FigS2}(a). The optimization iteratively updates the unknown parameters to minimize the reduced chi-squared error and terminates when the change between successive iterations falls below a threshold denoted by $\epsilon$, with $\epsilon$ set to $10^{-3}$. Multiple initial guesses of the Drude-Lorentz parameters were tested to ensure convergence to the global minimum.}

\highlightblue{To reduce the unknowns in the material stack to only the refractive indices of the deposited thin films, the thickness and refractive index of the substrate needed to be known. The \si{CaF_2} substrate used in this work was a double-sided polished, 500 $\mu$m-thick, 10$\times$10 \si{mm^2} sample with (100) crystal orientation, purchased from MTI Corp (SKU: CFa101005S2). Although several studies have reported the refractive indices of \si{CaF_2}, they did not provide the imaginary part below 10 $\mu$m. Li provided the real part from 0.15 to 12 $\mu$m wavelength using Sellmeier formula \cite{R43}, and Kaiser et al. reported a Lorentz oscillator model from 10 to 80 $\mu$m wavelength \cite{R2}. Therefore, we decided to measure the FTIR transmission and reflection spectra of bare \si{CaF_2} and extracted its refractive indices. A two-oscillator Lorentz model was used to fit the FTIR measurement range, but it could not simultaneously capture the near-zero extinction coefficient below 10 $\mu$m and the increased absorption above 20 $\mu$m. Hence, we employed a model-independent fitting approach by treating the complex refractive index at each wavelength as an independent variable, \( n(\lambda) + i\,k(\lambda) = x_1 + i\,x_2 \), and optimizing the values of $x_1$ and $x_2$ at each wavelength using the FTIR transmission and reflection data. To ensure continuity and suppress unphysical fluctuations between neighboring wavelength points, the extracted complex refractive index of \si{CaF_2} was post-processed using MATLAB’s \texttt{smoothdata} function. The resulting mean values, along with uncertainty bounds obtained from the propagated uncertainty in the Monte Carlo simulations, are shown in Figure \ref{fig:FigS2}(b).} 

\highlightblue{The extracted refractive indices of \si{CaF_2} were then used in the dielectric extraction function for NbN and MoSi to calculate their refractive indices from the measured FTIR spectra. We used one Drude and one Lorentz oscillator for NbN, and one Drude and two Lorentz oscillators for MoSi, as detailed in Table I. The final reduced chi-squared error ($\chi^2_{red}$) values were 1.139 for NbN and 1.206 for MoSi. The reduced chi-squared errors being slightly greater than one in both cases are likely due to underestimated measurement uncertainties in the incidence angle and substrate thickness, as these uncertainties were not accurately characterized. The simulated spectra, generated using the extracted indices and their associated uncertainty bounds, were compared with the FTIR measurements in Figures \ref{fig:FigS2}(c) and \ref{fig:FigS2}(d), respectively.}

\vspace{-0.8cm}
\subsection{Validation of Spectral Energy Balance}
\vspace{-0.5cm}
\highlightblue{The \si{CaF_2} substrate used in this work is 500 $\mu$m thick and highly transparent in the mid-infrared range. Hence, light incident on \si{CaF_2} at oblique angles can cause multiple internal reflections, and some light might not be captured by the FTIR reflectance accessories. To evaluate whether losses due to these reflections are significant in our system, we performed a spectral energy balance calculation over the wavelength range of 2.5 to 7.5 $\mu$m. We validated energy conservation by summing the transmittance and reflectance spectra of bare \si{CaF_2} over this wavelength range since the extinction coefficient of \si{CaF_2} is extremely low in this wavelength range, as shown in Figure \ref{fig:FigS2}(b).}
\begin{figure}[!t]
\vspace{-0.5cm}
\includegraphics[width=\linewidth]{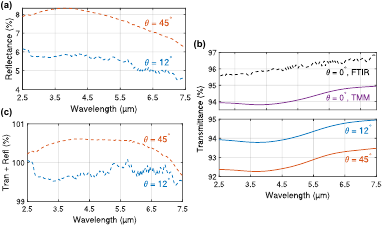}
\vspace{-1.35cm}
\caption{\footnotesize \protect\highlightblue{(a) FTIR reflectance spectra of a bare \si{CaF_2} at 12$^\circ$ and 45$^\circ$ incidence angles. (b) FTIR transmittance spectrum at normal incidence (dashed line) and simulated spectrum using the transfer matrix method (TMM) (solid line). TMM-based simulated transmittance spectra at 12$^\circ$ and 45$^\circ$ incidence angles are also shown. (c) Sum of transmittance and reflectance spectra of a bare \si{CaF_2} at 12$^\circ$ and 45$^\circ$ incidence angles.} \vspace{-0.25cm}}
\label{fig:FigS5}
\end{figure}

\highlightblue{Figure \ref{fig:FigS5}(a) shows the FTIR reflectance spectra of bare \si{CaF_2} at 12$^\circ$ and 45$^\circ$ incidence angles. The transmittance of \si{CaF_2} at normal incidence was simulated using the extracted refractive indices of \si{CaF_2} by applying the transfer matrix method and compared with the measured FTIR data, as shown in \ref{fig:FigS5}(b). Since the effect of multiple internal reflections is negligible at normal incidence, the observed discrepancy of approximately 1.5\% between the simulated and measured spectra is attributed primarily to uncertainties in the FTIR transmittance measurement, the \si{CaF_2} thickness measurement, and the propagated uncertainty in the extracted refractive indices of \si{CaF_2}. We then simulated the transmittance of \si{CaF_2} at 12$^\circ$ and 45$^\circ$ incidence angles using transfer matrix method. Experimental validation of the transmittance at these angles was not possible as we do not have accessories to measure transmittance at 12$^\circ$ and 45$^\circ$ incidence angles. However, since the transfer matrix method assumes that all multiple internal reflections are captured, these simulated spectra should not be affected by losses from multiple internal reflections and are instead influenced only by uncertainty in the extracted refractive indices, which is expected to be minimal.}

\highlightblue{Figure \ref{fig:FigS5}(c) shows the sum of the FTIR reflectance and simulated transmittance of the bare \si{CaF_2} substrate at 12$^\circ$ and 45$^\circ$ incidence angles, which is close to 100\%. Although the reflectance not captured by FTIR measurements due to multiple internal reflections may be compensated by an overestimation in the simulated transmittance caused by uncertainty in the extracted refractive indices of \si{CaF_2}, we expect this overestimation to be minimal and to remain within the typical FTIR measurement uncertainty range of $\sim 1\%$. Hence, we can consider that the effect of multiple internal reflections at oblique incidence is not significant in our system and is likely encompassed within the overall FTIR measurement uncertainty.}

\vspace{-0.7cm}
\subsection{Calculation of DC Resistivity from Drude Dielectric Coefficients}
\vspace{-0.5cm}
The dielectric functions of metals, conductive oxides, and heavily doped semiconductors are typically modeled using the Drude-Lorentz model. The Drude term accounts for the intraband transition of free electrons in the material, and it is characterized by two parameters: the plasma frequency ($\omega_p$) and a broadening factor related to the free electron lifetime ($\Gamma_D$). Following the equations in \cite{R31, R32}, the density of conduction electrons in a material can be calculated from its unscreened plasma energy ($\omega_p$) using $N = (\omega_p^2 \epsilon_0 m^*)/(e^2 \hbar^2)$. Here $\epsilon_0$ is the free-space permittivity, $m^*$ is the electron effective mass, $e$ is the electron charge, and $\hbar$ is the reduced Planck constant. The other Drude parameter, $\Gamma_D$, is associated with the scattering of conduction electrons in the material. According to the free-electron model, the relaxation time of the conduction electrons is given by $\tau = \hbar/\Gamma_D$ \cite{R32}. In addition, the relaxation time is related to the room-temperature resistivity ($\rho_n$) of the film by the equation $\rho_n = m^*/(\tau N e^2)$ \cite{R31,R32}. By substituting the values of $\tau$ and $N$ in this equation, we get
\begin{equation*}
\begin{aligned}
\rho_n = \frac{\Gamma_{D} \hbar}{\omega_p^2 \epsilon_0 e}
\end{aligned}
\end{equation*}
Thus, by obtaining the fitted Drude parameters, namely $\omega_p$ and $\Gamma_D$, for NbN and MoSi from our refractive index extraction algorithm, we can calculate the room-temperature resistivity values of the films.
\begin{figure}[!t]
\vspace{-0.5cm}
\includegraphics[width=0.6\linewidth]{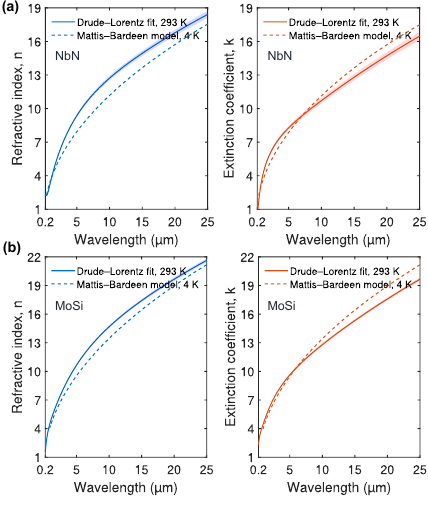}
\vspace{-0.5cm}
\caption{\footnotesize The refractive index and extinction coefficient of (a) 10-nm-thick NbN and (b) 15-nm-thick MoSi films are shown, extracted from room-temperature FTIR measurements (solid line: \protect\highlightblue{mean values, shaded region: uncertainty bounds}) and simulated using the Mattis-Bardeen model at 4 K (dashed line). \protect\highlightblue{In the simulation, the critical temperatures (\si{T_{\mathit{c}}}) of 10-nm-thick NbN and 15-nm-thick MoSi were taken as 9.67 K and 5.23 K, respectively.}} \vspace{-0.5cm}
\label{fig:FigS6}
\end{figure}

\vspace{-0.6cm}
\subsection{Refractive Indices of NbN and MoSi Thin Films Below \si{T_{\mathit{c}}}}
\vspace{-0.5cm}
In this section, we discuss the validity of using the room-temperature mid-infrared refractive indices of NbN and MoSi to represent their refractive indices at cryogenic temperatures. At zero temperature, an ideal superconductor does not absorb optical radiation with energies lower than its superconducting gap energy (2\si{\Delta_0}). However, when the photon energy is well above the superconducting gap energy, an individual quantum of electromagnetic radiation can break a Cooper pair and generate two excited quasiparticles. This leads to dissipation in the superconductor, and the optical conductivity of the superconducting film below its critical temperature (\si{T_{\mathit{c}}}) can be derived using the Mattis-Bardeen model \cite{R4,R7}:
\begin{equation}
\begin{aligned}
\epsilon_{rs}^{'}|_{\hbar \omega < 2\Delta} = 1 - \frac{\sigma_n}{\hbar \epsilon_0 \omega^2} \int_{\Delta-\hbar \omega}^{\Delta} [1-2f(E+\hbar \omega)] \times \frac{[E^2+\Delta^2+\hbar\omega E]}{[\Delta^2-E^2]^{1/2} [(E+\hbar\omega)^2-\Delta^2]^{1/2}} \, dE \\
\epsilon_{rs}^{'}|_{\hbar \omega > 2\Delta} = 1 - \frac{\sigma_n}{\hbar \epsilon_0 \omega^2} \int_{-\Delta}^{\Delta} [1-2f(E+\hbar \omega)] \times \frac{[E^2+\Delta^2+\hbar\omega E]}{[\Delta^2-E^2]^{1/2} [(E+\hbar\omega)^2-\Delta^2]^{1/2}} \, dE
\end{aligned}
\end{equation}

\begin{equation}
\epsilon_{rs}^{''} = \frac{2\sigma_n}{\hbar \epsilon_0 \omega^2} \int_{\Delta}^{\infty} [f(E)-f(E+\hbar \omega)]g(E) \, dE + \int_{\Delta-\hbar \omega}^{-\Delta} [1-2f(E+\hbar \omega)]g(E) \, dE
\end{equation}
here $\sigma_n$ is the normal state conductivity, $f(E) = 1/(1+\exp(E/k_BT))$ is the Fermi-Dirac function, and $g(E) = [E^2+\Delta^2+\hbar\omega E]/([E^2-\Delta^2]^{1/2} [(E+\hbar\omega)^2-\Delta^2]^{1/2})$ is the density of state. The term $2\Delta$ represents the superconducting gap energy at $T$ = 0 K. For temperatures $T<T_\mathit{c}$, we can find $\Delta = 1.74 \times \Delta_{T=0} \sqrt{1-(T/T_\mathit{c})}$. The value of $\sigma_n$ can be calculated from the inverse of the room-temperature resistivity ($\rho_n = R_{sheet} \times \text{thickness}$) of the film. \highlightblue{In this calculation, we considered the sheet resistance ($R_{sheet}$) and thickness of the NbN and MoSi films deposited on \si{CaF_2} substrates, as given in Table \ref{tab:S1}.}

Finally, the complex refractive index of the superconducting film is given by $n+ik= \sqrt{\epsilon_{rs}^{'}+i\epsilon_{rs}^{''}}$ where $n$ and $k$ are the refractive index and extinction coefficient, respectively. The Mattis-Bardeen model has been applied to different type-I and type-II BCS superconducting thin films previously and has shown good agreement with experimental results \cite{R35,R33,R34}. Figure \ref{fig:FigS6} shows the calculated refractive indices of NbN and MoSi below their critical temperatures using the Mattis-Bardeen model, compared with the values measured at room temperature via FTIR spectroscopy. The plot suggests that the optical properties of the superconductors at room temperature provide a good approximation of their properties at cryogenic temperatures. This aligns with the commonly held perception in the literature \cite{R11, R12}, as SNSPDs are often designed using the room-temperature refractive indices of superconducting materials \cite{R11,R12,R17}. Our calculation shows that this assumption holds true for the mid-infrared spectrum.

%